\documentclass[pra,twocolumn,aps,showpacs,superscriptaddress,floatfix,showpacs]{revtex4-1}
\usepackage{graphicx}
\usepackage{amsmath}
\usepackage{times}
\usepackage[usenames,dvipsnames]{color}
\usepackage[colorlinks,bookmarks=false,citecolor=blue,linkcolor=red,urlcolor=blue]{hyperref}
\usepackage{float}
\usepackage{amsfonts}

\begin{document}
\title{Quantum Dynamics of Solitons in Strongly Interacting Systems on Optical Lattices}
\date{\today}

\author{Chester P. Rubbo}
\affiliation{JILA, (NIST and University of Colorado), and Department of Physics, University of Colorado, Boulder, Colorado 80309-0440, USA.}

\author{Indubala I. Satija}
\affiliation{Department of Physics, George Mason University, Fairfax, VA 22030, USA.}
\affiliation{Joint Quantum Institute, National Institute of Standards and Technology and University of Maryland, Gaithersburg, MD 20899, USA.}

\author{William P. Reinhardt}
\affiliation{Joint Quantum Institute, National Institute of Standards and Technology and University of Maryland, Gaithersburg, MD 20899, USA.}
\affiliation{Department of Chemistry, University of Washington, Seattle, WA 98195-1700, USA.}

\author{Radha Balakrishnan}
\affiliation{The Institute of Mathematical Sciences, Chennai 600113, India}

\author{Ana Maria Rey}
\affiliation{JILA, (NIST and University of Colorado), and Department of Physics, University of Colorado, Boulder, Colorado 80309-0440, USA.}

\author{Salvatore R. Manmana}
\affiliation{JILA, (NIST and University of Colorado), and Department of Physics, University of Colorado, Boulder, Colorado 80309-0440, USA.}

\date{\today}
\begin{abstract}
Mean-field dynamics of strongly interacting bosons described by hard core bosons with nearest-neighbor attraction
has been shown to support two species of solitons: one of Gross-Pitaevskii-type (GP-type) where the condensate fraction remains dark and a novel
non-Gross-Pitaevskii-type (non-GP-type) characterized by brightening of the condensate fraction. 
Here we study the effects of quantum fluctuations on these solitons using the adaptive time-dependent density matrix renormalization group method, 
which takes into account the effect of strong correlations. 
We use local observables as the density, condensate density and correlation functions as well as the entanglement entropy to characterize the stability of the initial states.  
We find both species of solitons to be stable under quantum evolution for a finite duration, their tolerance to quantum fluctuations being enhanced as the width of the soliton increases. 
We describe possible experimental realizations in atomic Bose Einstein Condensates, polarized degenerate Fermi gases,  and in systems of polar molecules on optical lattices. 
\end{abstract}
\maketitle
\section{Introduction}
\label{sec:intro}

Solitary waves and solitons (i.e., solitary waves whose shape and speed remain unchanged even after collisions) are encountered in systems as diverse as classical water waves \cite{book}, magnetic materials 
\cite{book,OshikawaAffleck97,Dender,Essler99,AffleckOshikawa99prb_a,AffleckOshikawa99prb_b,Asano,Feyerherm,Kohgi,Wolter,EsslerFH,Kenzelmann,Zvyagin_solitons,Nojiri,Umegaki,NIST,Zvyagin2011}, fiber-optic communication \cite{book,book_opticalsolitons,Agrawal}, and Bose-Einstein condensates (BEC) \cite{rmp_dalfovo,book,gpe,BECsolitonbook}. 
Rooted in the nonlinearity of the system which balances dispersive effects, solitons are fascinating non-linear waves that encode collective behavior in the system. 
Intrinsically nonlinear in nature due to inter-particle interactions, and due to the high degree of control in the experiments, the BEC systems are a natural fertile ground for exploring solitons. 
In dilute atomic gaseous BECs which are simply described in terms of the properties of the non-linear Schr\"odinger equation, or 
Gross-Pitaevskii equation \cite{gpe} (GPE),  bright (density elevation) solitons exist for attractive interparticle interactions and dark (density notch) solitons in the repulsive case.  
These solutions are characterized not only by persistent density profiles, but also by characteristic modulations of the quantum phase across their profiles, which differs for the bright (attractive condensate) \cite{CBPRL} and the dark (repulsive condensate) cases \cite{Burger:1999p5198,Denschlag:2000p97,Bexpt}. 
However, we want to emphasize that, on general grounds, other systems with intrinsic nonlinearities should be 
able to realize solitons if the conditions are chosen appropriately.  

In this paper, we follow two goals: 
First, we want to describe the realization of solitons in lattice systems since interaction effects there play a more pronounced role than in the aforementioned systems of dilute atomic gases.  
In this way, we can systematically study the effect of interactions on the soliton dynamics in the broader framework of possible experimental observations in BEC, quantum degenerate Fermi gases, hard-core bosons, and polar molecules on optical lattices, as well as in certain condensed matter systems.
The common aspect of these various systems is that the soliton dynamics can be described in terms of a simple $S=1/2$ spin chain which, in turn, can be the effective model for a variety of situations, as the ones mentioned above. 
This description is footed on an extension of the standard GPE treatment of solitons, and leads us to the second scope of our paper which is to describe new effects which go beyond mean-field dynamics.

Investigations of effects beyond GPE dynamics have been a subject of various studies in the past decade. 
For short range repulsive systems, the cubic non-linearity of the
GPE was replaced by a quintic repulsive nonlinearity and the resulting modified GPE was shown to support dark solitary waves of GP-type \cite{Kolo}. 
This 1D system was further investigated in the presence of dipolar interactions \cite{Dipole} and was shown to support bright solitons whose stability and mobility depended on the dipolar interaction strength.  
In 2D systems, bright solitons were found to be stable given a sufficient dipole-dipole strength \cite{Pedri:2005p200404}.  Further studies of the stability and dynamics of solitons have been extended to two component BECs \cite{Ohberg:2000p2918} and multilayered BECs \cite{Nath:2007p013606}.  
Existence of dark and bright solitary waves was also shown numerically in systems describing multicomponent BECs \cite{BECmix}.  
In addition to the continuum systems, solitary waves have been extensively studied in systems described by a discrete non-linear Schr\"odinger equation \cite{DNLS,Ahufinger:2004p053604}, BECs in deep optical lattices and also in optical beams in wave guides. 

In recent studies \cite{PRL,Eugene}, solitary waves in a system of hard core bosons (HCB) described in terms of
hard core on-site repulsion and attractive nearest neighbor interaction,
 have been studied using mean field equations obtained from mapping the HCB system to an anisotropic $S=1/2$ Heisenberg spin system.    
The continuum limit of the lattice populated with HCB is described by a generalized GPE, which we will refer to as "HGPE" [See Eq.~(\ref{eq:hgpe})]
as it describes hard core bosons in mean field treatment.

In contrast to the GPE, HGPE contains both the normal and condensate density. 
This system describing strongly repulsive BEC was shown to support both dark and bright solitary waves, the existence of both species being rooted in the particle-hole symmetry in HCB systems. 
Unlike other studies, HGPE solitary waves are  obtained as an 
analytic solution which was shown to provide an almost exact solution of the equations of motion \cite{PRL}. 
These two species of solitons can be referred to as the GP-type and the non-GP-type as the former corresponds to a dark condensate fraction that dies at sound velocity while the latter is associated with brightening of the condensate and persists all the way up to sound velocity and transforms into a soliton train for supersonic velocities.
Recent numerical studies investigating collision properties of these nonlinear modes suggest that these solitary waves are in fact solitons \cite{UNpub}. 

An important question that we investigate here is whether these mean field solitons survive quantum fluctuations.
In previous work, the quantum dynamics of GP dark solitons in the superfluid regime of the Bose-Hubbard Hamiltonian has been studied numerically by Mishmash et al. \cite{MishmashPRL2009,Carr}. 
The main findings are that for weak interactions the dark soliton is stable on a time-scale of the order of $\sim20-40$ units of the hopping and afterwords decays due to two-particle scattering processes. 
The larger the on-site interaction, the stronger the scattering and the faster the decay of the solitons. 
In addition, these studies treated collisions between the solitary waves which confirm the soliton nature of the states on the time scales treated.  
These studies focus on the limit of small interactions. 
Here, we treat the strong coupling case and study the fate of the soliton solutions obtained in the HGPE framework. 
We do this by generalizing the Bose-Hubbard model of Refs.~\cite{MishmashPRL2009,Carr} to include on-site and nearest neighbor density-density interactions. 
As discussed in Ref.~\cite{PRL}, in the continuum limit this gives rise to the two distinct types of solitons mentioned above, which, as we shall see, are found to be stable in both the mean field approximation to the lattice dynamics of the system, as well as in the full quantum dynamics on the lattice.  

More specifically, we describe the exact quantum evolution of an initial mean field soliton solution on 1D lattice systems.
The soliton and the Hamiltonian driving the dynamics are thereby formulated in terms of a $S=1/2$ spin language.   
It is so possible to envisage a realization of the described soliton solutions in both, experiments with ultracold bosonic and spin polarized fermionic atoms, as well as in experiments with polar molecules \cite{carr09,ni08,aikawa10,deiglmayr08} on optical lattices which can be used to emulate spin-1/2 systems \cite{gorshkov1,gorshkov2}.   
We combine an analytic solution of the HGPE which provides a continuum approximation to the lattice problem, a numerical treatment of the mean-field equations on the lattice, and a full quantum treatment of the dynamics by applying the time-dependent DMRG \cite{white1992,white1993,Schollwock:2005p2117,Daley:2004p2943,White:2004p2941}. 
Both, mean-field and numerical results indicate that for a certain range of parameters the solutions found are indeed stable 
solitons on the time scale treated. 
The non-GP-type soliton 
is found to be somewhat less tolerant of quantum effects compared to the GP-type. 
We characterize the stability of the solitons by considering the entanglement in the system: since in our set-up the initial soliton solutions are product states on the lattice, the entanglement entropy \cite{Amico} should remain zero for a stable soliton solution and hence is a measure for the stability of the soliton in the course of the time evolution. 
In addition, we consider correlation functions which, on similar grounds, can be used to characterize its stability. 

The paper is organized as follows. 
In section~\ref{sec:model}, we introduce the effective spin model and its derivation from HCB and spinless fermions on a lattice, the dynamical equation (HGPE) that describes the continuum approximation to the mean-field equations of the lattice system, and we summarize the analytic solution of the HGPE. 
In Sec.~\ref{sec:methods} we describe the mean-field ansatz and some details of the DMRG approach to the dynamics.  
In Sec.~\ref{sec:numerics}, we analyze the stability of the soliton solutions by comparing the mean-field results on a lattice to the DMRG results.
As a measure for the quality of the soliton solution, we use in Sec.~\ref{sec:entropy} the von Neumann or entanglement entropy as well as correlation functions which also should remain zero in the course of the time evolution if the mean field state were to survive quantum fluctuations. 
In Sec.~\ref{sec:experiments} we propose possible experimental realizations of the HGPE solitons.   
In Sec.~\ref{sec:summary}, we summarize.  

\section{Hamiltonian and Equations of Motion}
\label{sec:model}

In this paper, we treat the dynamics of initial soliton states driven by the spin Hamiltonian
\begin{equation}
H_S = - \sum_j \left[ J \,\, \hat{\mathbf{S}}_j \cdot \hat{\mathbf{S}}_{j+1} - g\,\, \hat{S}_j^z \hat{S}_{j+1}^z \right] - g \sum_j{\textstyle}\,\, \hat{S}_j^z    
\label{eq:spinsystem}
\end{equation}
on a one-dimensional lattice, i.e., we are treating the dynamics of a XXZ-chain with a global external magnetic field of magnitude $g$.  
One way to obtain this effective Hamiltonian is as the limiting case of the extended Bose Hubbard model, 
\begin{equation}
\begin{split}
H = &-\sum_j \left[ \frac{J}{2} \, \left[ b_j^{\dagger} b_{j+1}^{\phantom{\dagger}} + h.c. \right]+ V n_j^{\phantom{\dagger}} n_{j+1}^{\phantom{\dagger}} \right] \\
&+  \sum_j \left[ \frac{U}{2} n_{j}^{\phantom{\dagger}} \left(n_{j}^{\phantom{\dagger}} - 1 \right) 
-\left( \mu - J \right) n_{j}^{\phantom{\dagger}} \right]. 
\end{split}
\label {eq:BH}
\end{equation}
Here,  $b_j^{(\dagger)}$ are the annihilation (creation) operators for a  boson at the lattice site $j$, $n_{j}$ is the number operator, 
$J/2$ is the n.n. tunneling strength, and $\mu$ is the chemical potential. 
An attractive nearest-neighbor interaction $V < 0$ is introduced to soften the effect of a strong onsite interaction $|U| \gg 0$.  
The HCB limit  ($|U| \rightarrow \infty$) corresponds to the constraint that two bosons cannot occupy the same site. 
This HCB system can then be mapped to the model Eq.~(\ref{eq:spinsystem}) \cite{PTP.16.569}, where the two spin states correspond to two allowed boson number states $|0\rangle$ and $|1\rangle$, and  
setting $g=J-V$.  
Note that this is in contrast to the study of Refs.~\cite{MishmashPRL2009,Carr} in which the quantum dynamics was investigated in the original Bose-Hubbard system and not for the effective model Eq.~(\ref{eq:spinsystem}).  
This is interesting since the existence of the proposed soliton solutions for this spin model has implications for further systems than the ultracold bosonic atoms usually considered when describing soliton phenomena in cold gases.  
In particular it should be noted that the XXZ model in 1D can be obtained using the Jordan-Wigner transform from a system of spinless fermions
\begin{equation}
H_{\rm SF} = -\frac{J}{2} \sum\limits_j \left[ c_{j+1}^\dagger c_j^{\phantom{\dagger}} + h.c. \right] + V \sum\limits_j n_j^{\phantom{\dagger}} n_{j+1}^{\phantom{\dagger}} + \tilde{\mu} \sum_j n_j^{\phantom{\dagger}}, 
\label{eq:sfmodel}
\end{equation}
with $c_j^{(\dagger)}$ the fermionic annihilation (creation) operators on site $j$, and $n_j = c_j^\dagger c_j^{\phantom{\dagger}}$ the density on site $j$.  
Therefore, it should be possible to investigate the soliton dynamics in experiments with bosonic atoms, in spin systems, and in fermionic systems. 
In Sec.~\ref{sec:experiments} we discuss possible implementations in experiments with cold gases. 
Note that both models, Eq.~(\ref{eq:spinsystem}) and Eq.~(\ref{eq:sfmodel}) are fundamental models for describing condensed matter systems such as quantum magnets and systems of itinerant electrons. 
It is therefore conceivable that, in principle, the proposed soliton solutions can be realized in such systems as well. 
 
For simplicity, we set up our discussion in the framework of bosonic systems, without losing generality.  
Then, the spin flip operators $\hat{S}^{\pm}=\hat{S}_x\pm i\hat{S}_y$  correspond to the annihilation and creation operators of the corresponding bosonic Hamiltonian, $b_j \rightarrow \hat{S}_{j}^{+}$. 
Thus the order parameter that describes a BEC wave function is $\psi_j^s=\langle S_j^+\rangle$,  where the expectation value is obtained using spin coherent states \cite{Radha}.  In this mean-field description, the evolution equation for the order parameter is obtained by taking the spin-coherent state average of the Heisenberg  equation of motion.   
The spin coherent state $|\tau_j \rangle$ at each site $j$ can be parametrized as:
\begin{equation}
|\tau_j \rangle = e^{ i\frac{\phi_j}{2}} \left[ e^{-i\frac{\phi_j}{2}} \cos \frac{\theta_j}{2} |\uparrow \rangle + e^{i \frac{\phi_j}{2}} \sin \frac{\theta_j}{2} |\downarrow \rangle \right].
\label{eq:coh}
\end{equation}
With this choice, the HCB system is mapped to a system of classical spins \cite{PRL,Radha} via 
$\bf{S}= \left( \frac{1}{2}\sin(\theta) \cos(\phi),\frac{1}{2} \sin(\theta) \sin(\phi),\frac{1}{2} \cos(\theta) \right)$. 
Note that the particle density $\rho_j$ and the condensate density $\rho_j^s$ satisfy the relation $\rho^s_j= \rho_j \, \rho^h_j$, with $\rho^h_j = 1-\rho_j$ the hole density.  
In this representation,
$\psi_j^s = \sqrt{\rho_j^s} e^{i \phi}$.  This mean field treatment is contrasted to the standard GPE derived from the Bose-Hubbard model by taking the expectation value of the Heisenberg equation of motion with Glauber coherent states \cite{Langer}.  
We cast the equations of motion in terms of the canonical variables $\phi_j$ and $\delta_j \equiv \cos(\theta_j)=(1-2\rho_j)$ and obtain 
\begin{eqnarray}
\dot{\delta_j} &=& \frac{J}{2} \sum_{i = \pm 1} \sqrt{(1-\delta^2_j)(1-\delta^2_{j+i})} \sin (\phi_{j+i}-\phi_j) \nonumber\\
\dot{\phi_j} &=& \frac{J}{2} \frac{\delta_j}{\sqrt{(1-\delta^2_j)}} \sum_{i = \pm 1} \sqrt{1-\delta^2_{j+i}}\cos(\phi_{j+i}-\phi_j) \nonumber \\
&& - \frac{V}{2}\sum_{i = \pm 1} \delta_{j+i} - (J-V)\delta_0 \quad.
\label{eq:canonical}
\end{eqnarray}

\subsection{Solitary Waves in the Continuum Approximation}
\label{sec:continuum}

In the continuum approximation, the equations for the order parameter are derived from a Taylor series in the lattice spacing $a$ \cite{Radha2},
\begin{equation}
i \hbar \dot{\psi^s} = -\frac{\hbar^2}{2m} (1-2\rho) \nabla^2 \psi^s -V_e \psi^s \nabla^2 \rho +U_e\rho \psi^s -\mu \psi^s 
\label{eq:hgpe}
\end{equation}
where $J a^2 =\frac{\hbar^2}{m}$,
$U_e = 2(J-V)$ and
$V_e =V a^2$.
This equation can be viewed as a generalized-GPE and we will refer to it as the HGPE in view of its relation to HCBs.
The corresponding discrete Eq.~(\ref{eq:canonical}) will be referred to as discrete HGPE.  These equations have been shown to support solitary waves \cite{PRL} riding upon
a background density $\rho_0$: $\rho(z)=\rho_0+f(z)$, with $z=x-vt$.
We obtain for the soliton solution 
\begin{equation}
f(z, \rho_0)^{\pm}=\frac {2\gamma^2 \rho_0\rho_0^h}{\pm \sqrt{(\rho_0^h-\rho_0)^2+4\gamma^2 \rho_0 \rho_0^h} \; 
\cosh \frac{z}{\Gamma}-(\rho_0^h-\rho_0)},
\label{eq:feqn}
\end{equation}
where $\gamma=\sqrt{1-\bar{v}^2}$, and $\bar{v}$ being the speed of the solitary wave in units of $c_s=\sqrt{2\rho_0^s(1-V/J)}$, which is the speed of sound of the Bose gas system determined from its Bogoliubov spectrum \cite{Radha}.
$\Gamma$ is the width of the soliton,
\begin{equation}
\Gamma^{-1}=\gamma \sqrt{{\frac{2(1-\frac{V}{J})\rho_0 \rho_0^h}{\frac{1}{4}(\rho_0^h-\rho_0)^2+\frac{V}{J}\rho_0 \rho_0^h}}}. 
\label{eq:width}
\end{equation}
The characteristic phase jump associated with the solitary waves is 
\begin{equation}
\Delta \phi_{\pm}= (\sqrt{1-2c_s^2}) \cos^{-1} \frac{\bar{v}(1-2\rho_0)}{1-2\rho^s_0 \bar{v}^2}. 
\label{eq:phasejump}
\end{equation}

\begin{figure}[t]
\includegraphics[width=0.43\textwidth]{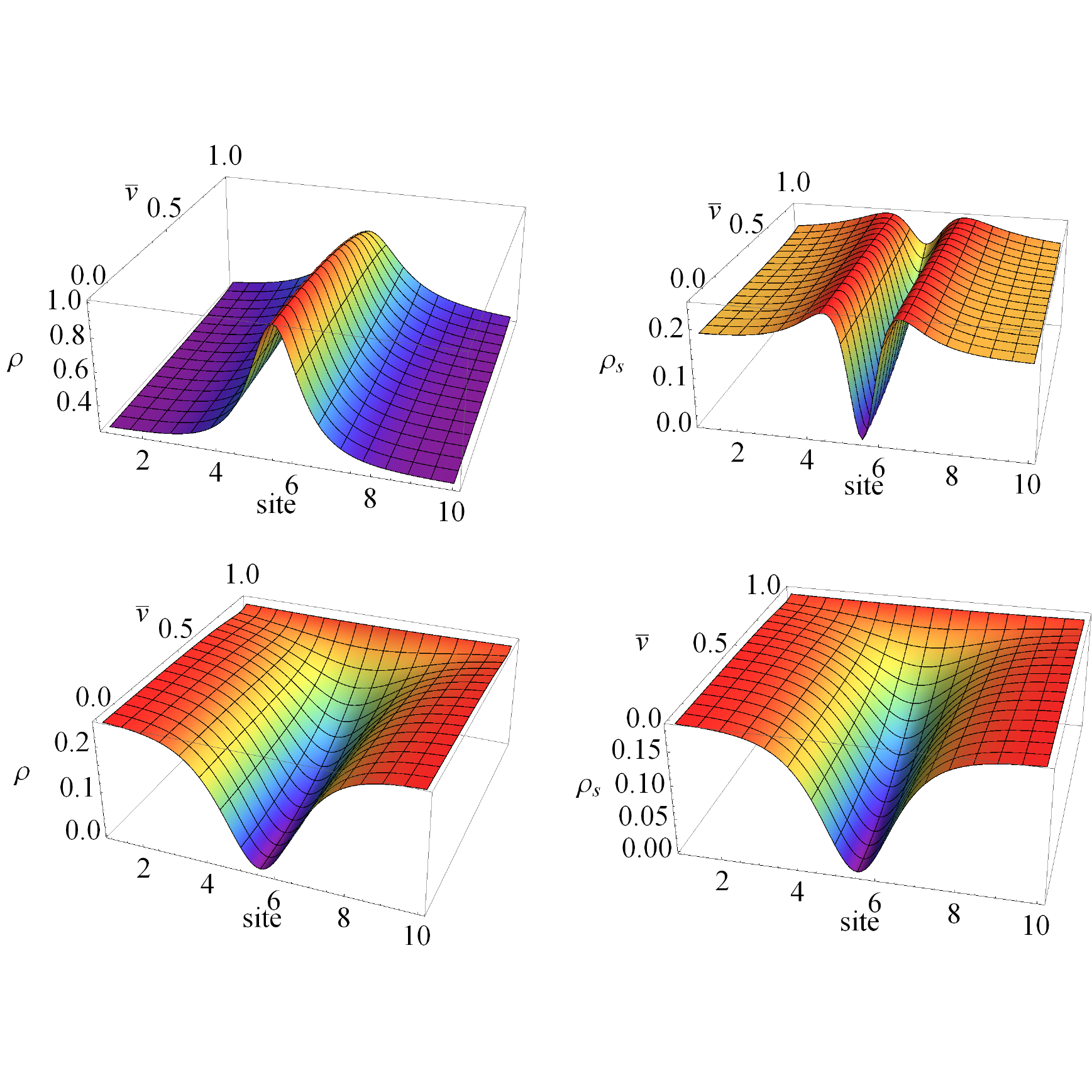}
\caption{(Color online)  Bright (top) and dark (bottom) soliton solution in the continuum [Eq.~\eqref{eq:feqn}] for $\rho_0 = 0.25$ and $V/J = 0.4$ .  Left panels show the density, right panels the condensate density as a function of position and speed. Note that the condensate density of the bright soliton shows a 'brightening', around the notch, i.e., it grows above the background value, whereas the dark soliton does not show this effect.} 
\label{fig:solitons}
\end{figure}
This solution has some remarkable properties. 
One direct consequence of the particle-hole symmetry underlying the equations of motion is the presence of two species of solitary waves, shown in Fig.~\ref{fig:solitons}.  
The existence of $f(z,\rho_0)$ superposed on the background particle density $\rho_0$  implies the existence of a counterpart $f(x, \rho_0^h)$, superposed upon a corresponding hole density $\rho_0^h$. 
In fact it is easy to see that $f^{\pm}(z,\rho_0) = \pm f^{\mp}(z,\rho_0^h)$. 
For $\rho_0 < 1/2$, the $\pm$ corresponds to bright and dark solitons, respectively. 
The bright solitons have the unusual property of persisting at speeds up to the speed of sound, in sharp contrast to the dark species that resembles the dark soliton of the GPE whose amplitude goes to zero at sound velocity. 
In the special case with background density equal to $1/2$, the two species of solitons become mirror images of each other, 
as $f^{+}(z,\rho_0=1/2) = - f^{-}(z,\rho_0=1/2)$. In this case, the condensate density in fact describes the GPE-soliton \cite{IRphyslett,pramana}. 

It should be noted that for $\rho_0 > 1/2$, the dark and bright solitons switch their roles. 
In other words, for $\rho_0 < 1/2$, it is the dark soliton that behaves like a GPE soliton while the bright soliton is the new type of soliton that persists all the way up to sound velocity. 
In contrast, for $\rho_0 > 1/2$, the bright soliton is GP-type while the dark one is the persistent soliton. 
In view of the particle-hole duality, we will present our results for $\rho_0 < 1/2$ in which the bright solitons have the persistent character noted above. 

In the following sections, we will investigate for the existence and the lifetime of these solutions on lattice systems using mean-field equations and the time dependent DMRG. 
We will complement this analysis by investigating the stability of further initial states.
In particular, we show that an initial Gaussian density distribution for a stationary soliton shows a similar stability if a phase jump is realized, but becomes unstable without a phase jump. 
This is of importance for experimental realizations indicating that imperfections in the creation of the initial state may not have a strong influence on the soliton dynamics. 

\section{Methods: mean field ansatz and DMRG}
\label{sec:methods}

\subsection{Mean field treatment}
\label{sec:meanfield}
\begin{figure}[t]
\includegraphics[width=.48\textwidth]{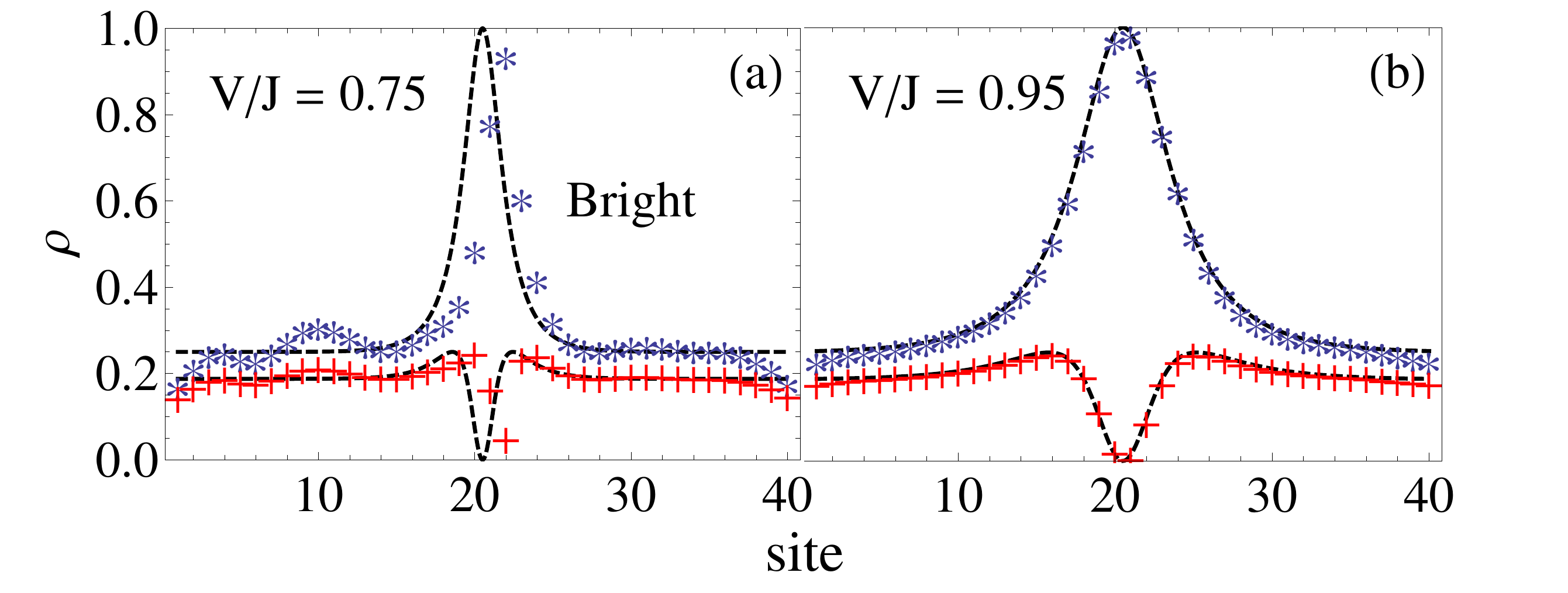}\\ 
\includegraphics[width=.48\textwidth]{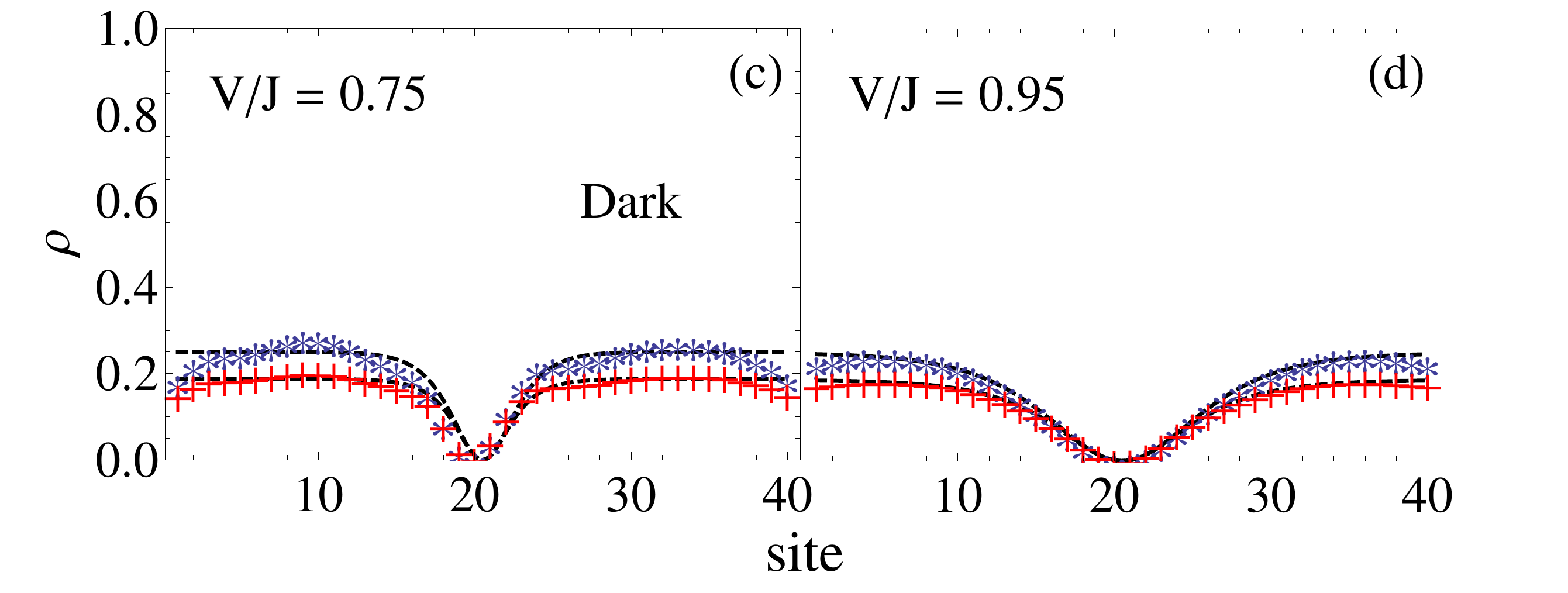} 
\caption{(Color online) 
Comparison of the soliton profiles for a background density $\rho_0 = 0.25$ at times $t = 20/J$ obtained in the continuum [black dashed line, Eq.~\eqref{eq:hgpe}] and using the equations of motion approach Eq.~\eqref{eq:canonical} on a lattice of $L=40$ sites. The left panels show the results for $V/J = 0.75$ (narrow soliton), the right panels the case $V/J = 0.95$ (broad soliton). The top panels show the bright soliton, the bottom panels the dark soliton solution of Eq.~\eqref{eq:feqn}. Both, the particle density $\rho$ $(*)$ and the condensate density $\rho^s$ $(+)$ are shown.} 
\label{fig:MeanFieldComparisons}
\end{figure}
  
\begin{figure}[b]   
\includegraphics[width=.48\textwidth]{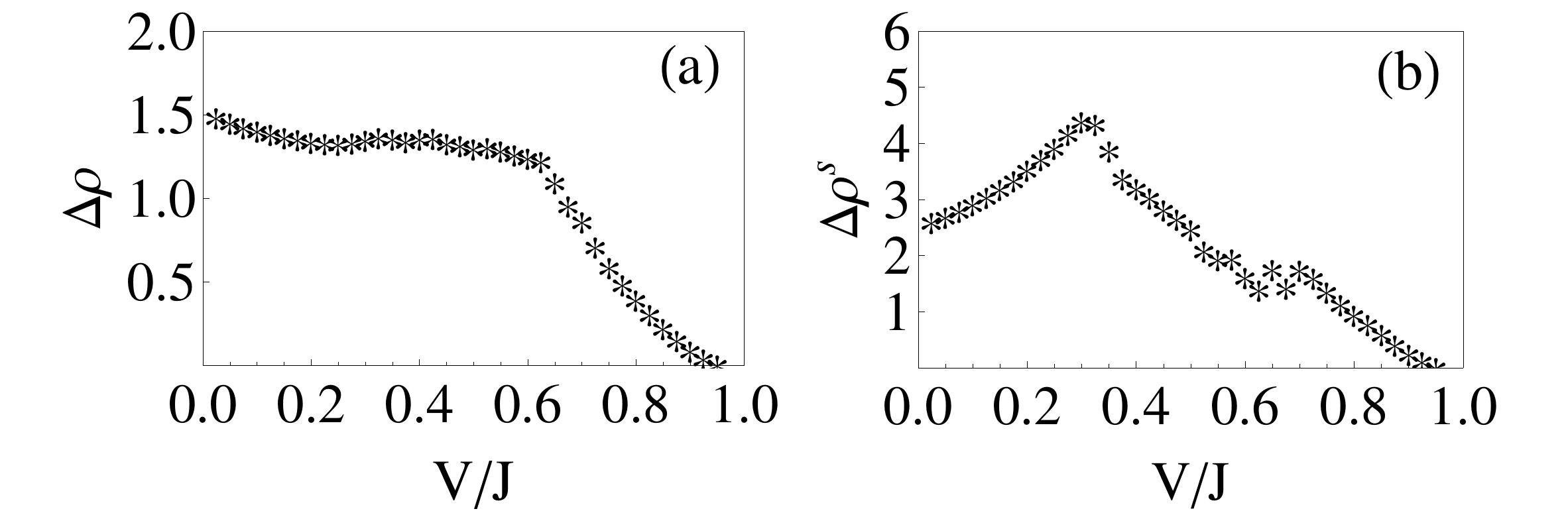} 
\includegraphics[width=.51\textwidth]{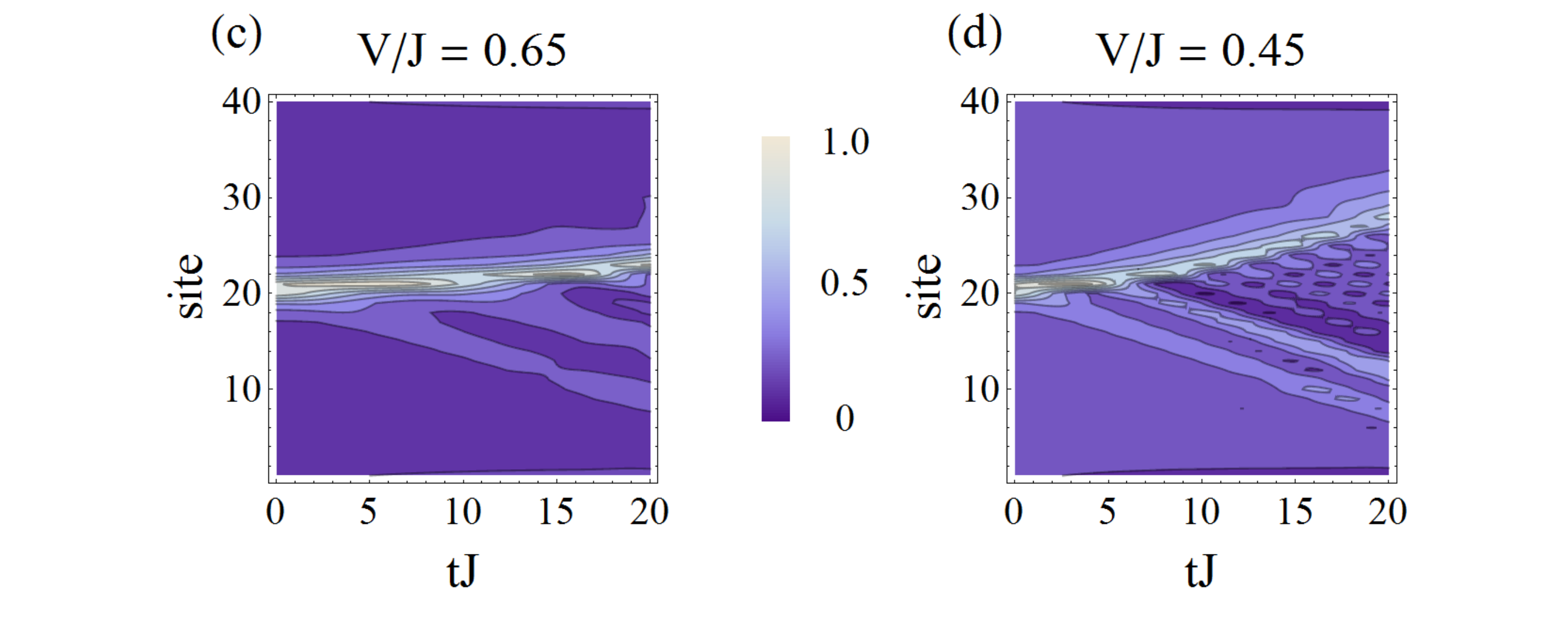}
\caption{(Color online) The differences resulting from Eq.~\eqref{eq:difference} of (a) the local density $\rho$ and (b) the condensate density $\rho^s$ between the mean-field continuum evolution and the mean-field lattice evolution on a lattice of $L=40$ sites at times $t = 20/J$ as a function of $V/J$. (c) and (d): lattice mean-field evolution of the local density for a broad soliton (c) with $V/J = 0.65$ and a narrow soliton (d) with $V/J = 0.45$.}
\label{fig:compare_meanfield_lattice}
\end{figure}
In this section, we compare the mean-field treatment of the soliton dynamics on a lattice [governed by Eqs.~\eqref{eq:canonical}] to the dynamics in the continuum [Eq.~\eqref{eq:hgpe}].  For the soliton dynamics on a lattice, we apply the equations of motion \eqref{eq:canonical} to an initial state given by Eqs.~\eqref{eq:feqn} and \eqref{eq:phasejump} on a finite lattice.  In Fig.~\ref{fig:MeanFieldComparisons} we show the discrete HGPE solitons for different values of $V/J$ at time $t = 20/J$. 
We compare the continuum solution (black dashed line) to the solution obtained on the lattice (symbols). 
As can be seen, for $V/J = 0.95$, the lattice approximation and the continuum solution show excellent agreement, up to small deviations at the boundaries.  For $V/J = 0.75$, however, significant deviations occur. 
We further analyze this behavior in Fig.~\ref{fig:compare_meanfield_lattice}
where we compute the difference of the lattice solution to the continuum solution in the local observables (density $\rho$ and condensate density $\rho^s$, respectively),
\begin{equation}
\Delta \rho^{(s)} = \frac{\sqrt{\displaystyle\sum_i \left( \left\langle \rho_i^{(s)} \right\rangle_{\rm continuum} - \left\langle \rho_i^{(s)} \right\rangle_{\rm MF} \right)^2}}{\sqrt{\displaystyle\sum_i  \left\langle \rho_i^{(s)} - \rho_0^{(s)}\right\rangle_{\rm continuum}^2}}
\label{eq:difference}
\end{equation}
at $t=20/J$ for a system of $L=40$ sites as a function of $V/J$.
\color{black}
As can be seen, the difference is significant for all values of $V/J \lesssim 0.8$.
Only at larger values the difference is of the order of a few percent. 

This discrepancy between the lattice and the continuum solution is to be expected: the continuum model is an approximation to the lattice model and its validity will break down when the size of features (e.g., the width of the soliton) of the analytic continuum solutions becomes comparable to the lattice spacings. 
This breakdown of validity can be understood in terms of the emission of Bogoliubov quasi-particles \cite{Pethickbook} for solitons which are too narrow: analogous to the excitations in a dilute bose gas, the Bogoliubov dispersion spectrum \cite{Radha} shows that a narrow perturbation excites high energy modes.
We further analyze this in Figs.~\ref{fig:compare_meanfield_lattice} (c) and (d). 
Quasi-particles are emitted in the course of the time evolution, and due to momentum conservation the soliton gets a velocity in the opposite direction so that it starts to move away from the original position.  
The narrower the soliton, the stronger the emission of quasi-particles, and -- as expected -- the lattice approximation becomes more and more unstable as the width of the soliton decreases, i.e., with decreasing the value of $V/J$. 

Note that this behavior is reminiscent of the mechanism which leads to the 'light-cone' effect in correlation functions following a quantum quench \cite{CalabresePRL2006,CalabreseJstat1,CalabreseJstat2,LauchliJstat,SRMlightcone,Naturelightcone}. 
In this case, the quench creates entangled quasi-particles on each lattice site which then move ballistically through the system and lead to a linear signature in the time evolution of correlation functions. 
In this way, the velocity of the quasi-particle excitations can be obtained \cite{LauchliJstat,SRMlightcone,Naturelightcone}. 
In a similar way, we propose that the linear signatures in Fig.~\ref{fig:compare_meanfield_lattice} can be used to further analyze the properties of the quasi-particles. However, this lies beyond the scope of the present paper so that we leave this issue open for future investigations.

Due to the necessity of having a width of the soliton larger than a few lattice spacings, we find that we need to investigate systems with $L \geq 30$ lattice sites.  Since this cannot be achieved using exact diagonalization methods for the Hamiltonian matrix, we choose to apply the adaptive t-DMRG which is capable of treating sufficiently large systems efficiently. 
In the following we therefore compare the lattice mean-field solution to the full quantum dynamics obtained by the DMRG for systems with $L = 40$ and $L=100$ lattice sites and $V/J \geq 0.9$.

\subsection{Details for the DMRG}
\label{sec:DMRG}

We apply the adaptive time-dependent extension of the density matrix renormalization group method \cite{white1992,white1993,Schollwock:2005p2117} (adaptive t-DMRG, \cite{Daley:2004p2943,White:2004p2941}) for systems with up to $L=100$ lattice sites with open boundary conditions. 
The DMRG is a numerical method which is capable of obtaining ground-state properties of (quasi-)one-dimensional systems with a very high efficiency and accuracy for lattices with up to several thousand sites, i.e., system sizes which are far larger than the ones amenable to exact diagonalizations of the Hamiltonian matrix. This is achieved by working in a truncated basis of eigenstates of reduced density matrices obtained for different bipartitions of the lattice. 
A measure for the error is given by the so-called discarded weight which is the sum of the weights of the density-matrix eigenstates which are neglected and which should be as small as possible (for more details, see, e.g., the review article \onlinecite{Schollwock:2005p2117}). 
Also its time-dependent extension can treat the real time evolution of strongly correlated quantum many-body systems substantially larger than the ones amenable to exact diagonalization methods and with an accuracy which can be, at short and intermediate times, similar to the one of ground state computations. In this paper, we exploit this accuracy in order to provide very high precision numerical results to which we compare the mean-field solutions discussed in Sec.~\ref{sec:meanfield}.  

\begin{figure}[t]
\includegraphics[width=.48\textwidth]{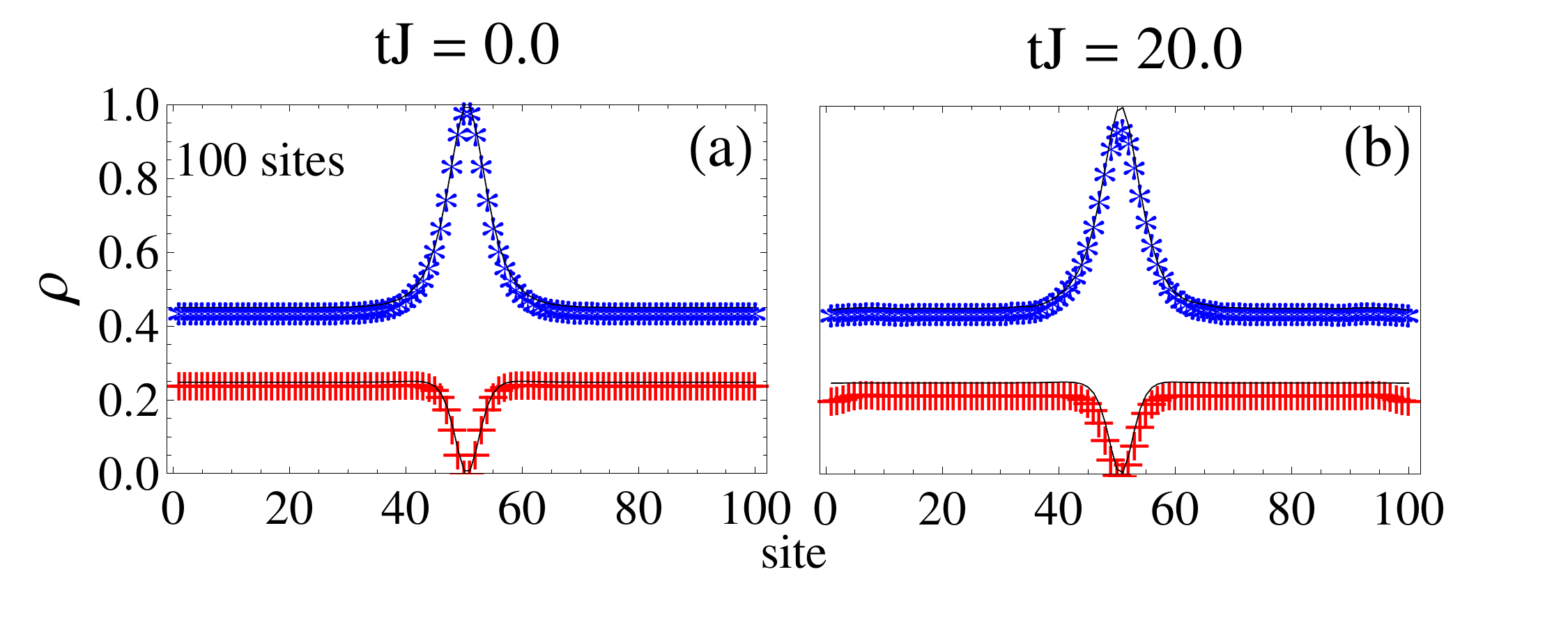} \\
\includegraphics[width=.48\textwidth]{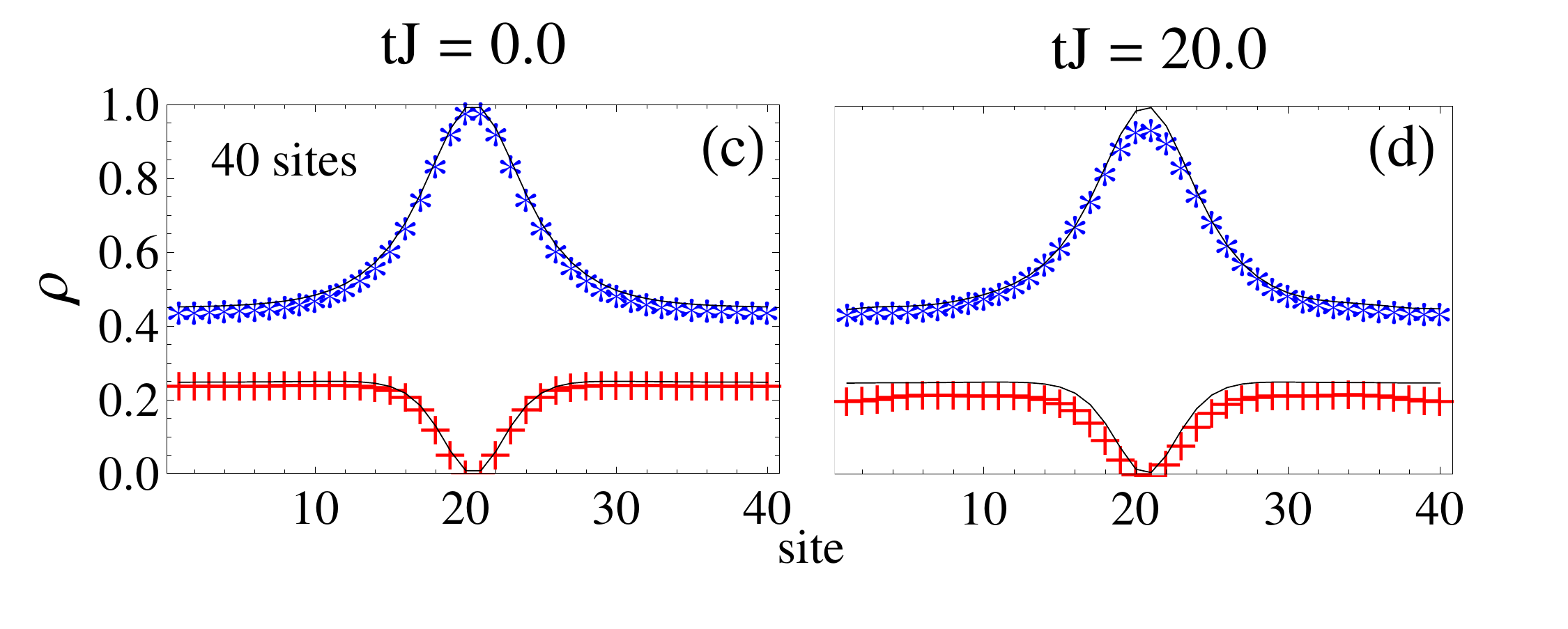}
\caption{(Color online) Local particle density $\rho$ (blue $*$) and condensate density $\rho^s$ (red $+$) obtained by DMRG (symbols) and by the mean-field ansatz (solid line) for a stationary bright soliton ($\overline{v} = 0$) at times $t=0$ (left panels) and at times $t=20/J$ (right panels). The plots show results for lattice sizes of $L=100$ sites (top) and $L=40$ sites (bottom). The parameters are $V/J = 0.95$ and $\rho_0 = 0.45$.}
\label{fig:Soliton40vs100}
\end{figure}

We solely use open boundary conditions since the DMRG performs far better in this case than in the case of periodic boundary conditions, so that we can treat larger system sizes with up to the aforementioned $L=100$ lattice sites.
However, at this point it becomes necessary to discuss the effect of the boundaries: we choose system sizes and initial widths of the solitons so that there is a wide region between the soliton and the boundary which can be considered to be 'empty'. 
In Fig.~\ref{fig:Soliton40vs100}, we compare the initial state for a system with $L=40$ and $L=100$ sites. 
As can be seen, the effect of the boundaries on the soliton is completely negligible. 
This remains so on time scales on which perturbations either from the boundaries reach the soliton or from the soliton reach the boundaries. 
At these instants of time, we stop the evolution and consider this to be the maximal reachable time for the given system size. 
We find that already for systems as small as $L=40$ sites, the maximal reachable time is $t > 20/J$, so that we conclude that the analysis which we present in the following is not affected by boundary effects. 

We work with the $S=1/2$ spin system [Eq.~(\ref{eq:spinsystem})] and engineer the initial state on the lattice by imprinting a phase and density profile by applying an external magnetic field.     
More specifically, for the initial state we compute the ground state of
\begin{equation}
H_0 = - h \sum_j \vec{B}_j \cdot \vec{S}_j
\end{equation}
with $h$ a large multiplicative factor ($\sim 100$) and 
\begin{eqnarray}
\vec{B}_j &=& \left\{ \langle S_j^x \rangle = \sqrt{\rho_j (1 - \rho_j)} \cos \phi_j, \right. \nonumber \\
               &&            \langle S_j^y \rangle = \sqrt{\rho_j (1 - \rho_j)} \sin \phi_j, \nonumber \\
               &&   \left.   \langle S_j^z \rangle = 0.5 - \rho_j \right\}. 
\end{eqnarray}
For the treatment of the dynamics of the system after turning off this magnetic field we apply a Krylov-space variant \cite{Manmana:2005p63} of the adaptive t-DMRG.   
During the evolution, we keep up to $1000$ density-matrix eigenstates for systems with up to $L=100$ sites.  
We apply a time step of $\Delta t = 0.05$, resulting in a discarded weight of $<10^{-9}$ at the end of the time evolution.  We estimate the error bars at the end of the time evolution to be smaller than the size of the symbols.   


\section{Full Quantum Dynamics}
\label{sec:numerics}

\begin{figure}[b]
\includegraphics[width=.48\textwidth]{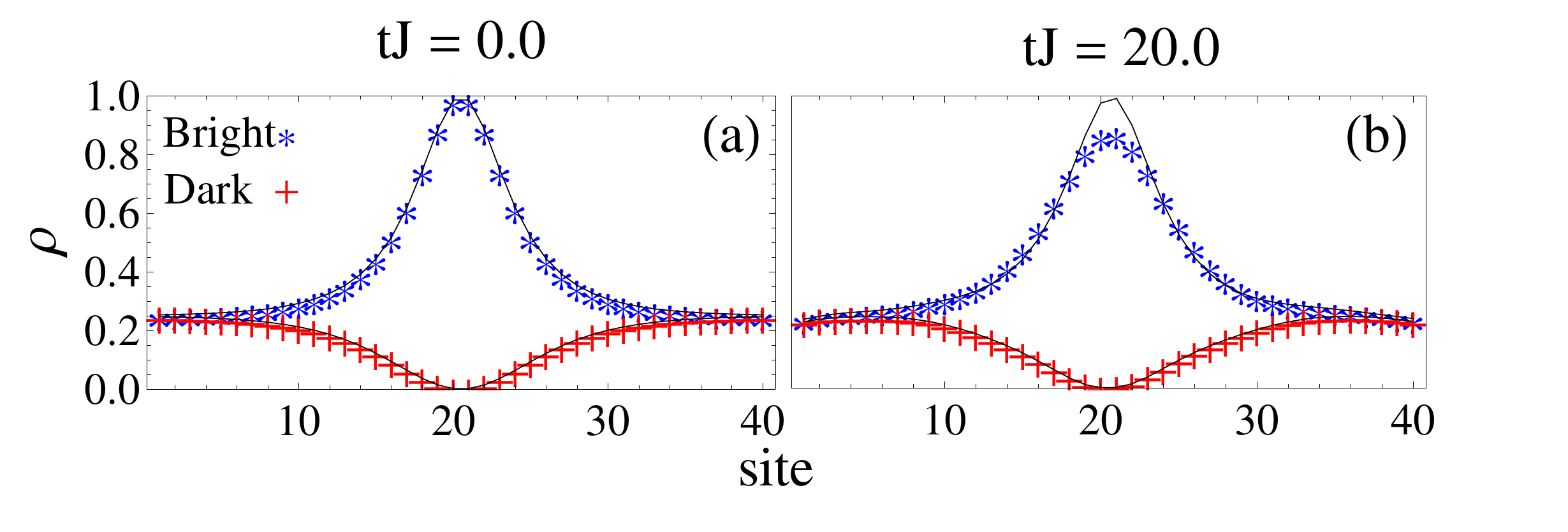} \\
\includegraphics[width=.48\textwidth]{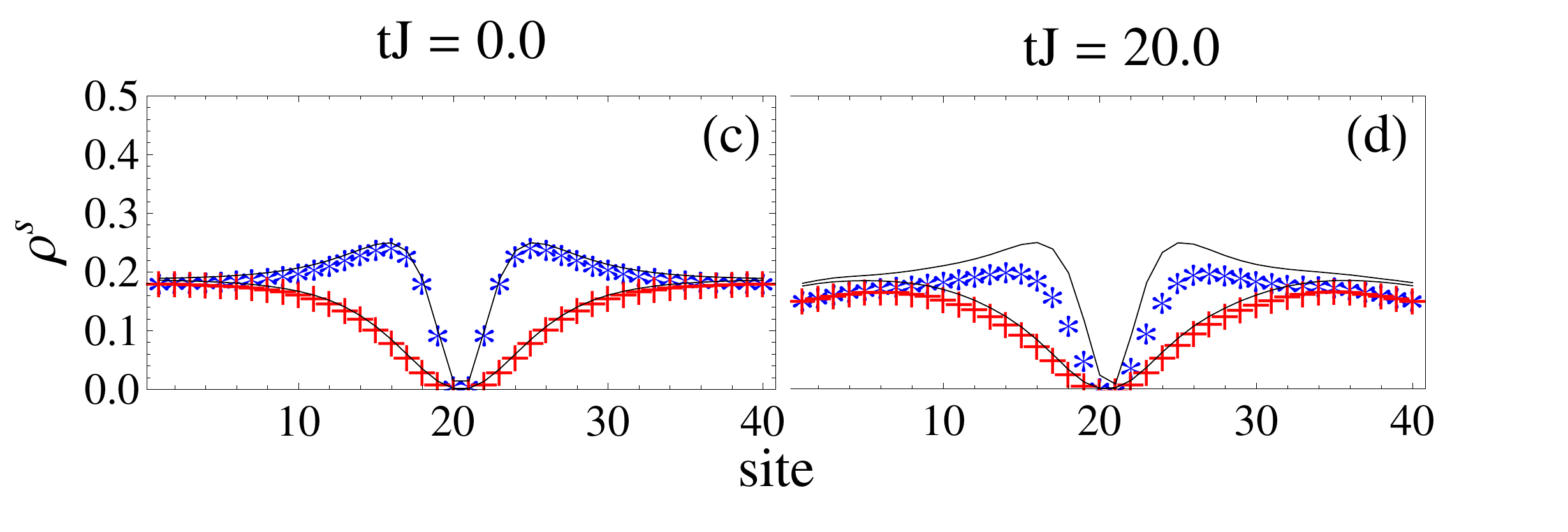}
\caption{(Color online) Local particle density $\rho$ (top) and condensate density $\rho^s$ (bottom) obtained by the DMRG (symbols) and by the mean-field (solid line) propagation of the stationary ($\overline{v} = 0$) bright (red $+$) and dark (blue $*$) solitons for times $t =  0$ and $20/J$ for a system with $L=40$ sites. The parameters are $V/J = 0.95$ and $\rho_0 = 0.25$.}
\label{fig:fig1}
\end{figure}

\subsection{Soliton Stability}
The initial soliton state is prepared as discussed in Section~\ref{sec:DMRG} and is propagated with the XXZ spin-$1/2$ Hamiltonian Eq.~(\ref{eq:spinsystem}) using the adaptive t-DMRG. 
Snapshots of the resulting time evolution for the density profile of both, the bright and the dark soliton,  with speed $\overline{v}=0$ are shown in Fig.~\ref{fig:fig1}. 
Since our results for moving solitons ($\overline{v} > 0$) are similar, we restrict in the following to the case of static solitons. 
While the mean-field solution remains essentially unchanged in time, the full quantum evolution shows some deformation of the initial state: in the course of the evolution, the total density profile widens as the peak decreases. 
The amount of change depends on the parameters $V/J$ and $\overline{v}$, and is different for the bright and the dark soliton.  
However, as further discussed below, for $V/J$ close enough to unity the difference between the quantum solution and the initial state remains below a few percent on a time scale $t \sim 20/J$, where the hopping amplitude due to the mapping from the spin system is $J/2$.
This has to be compared to time scales reachable by experiments on optical lattices. 
For typical lattice depths in which a tight binding description is valid, the tunneling rate varies from $0.1-1$ kHz, while the typical time scale of the experiments is on the order of 1-100 milliseconds.  
We therefore conclude that the density profile suggests a stable soliton on the experimentally accessible time scale in the full quantum evolution. 
Now we turn to the condensate density. 
Here, at $t=20/J$, the deviation from the mean field solution is larger.   
Nevertheless, as shown in Fig.~\ref{fig:fig1}, the change remains within a few percent for $V/t = 0.95$, so that we conclude that both quantities identify a stable soliton solution on this time scale. 

\begin{figure}[t]
\includegraphics[width=.53\textwidth]{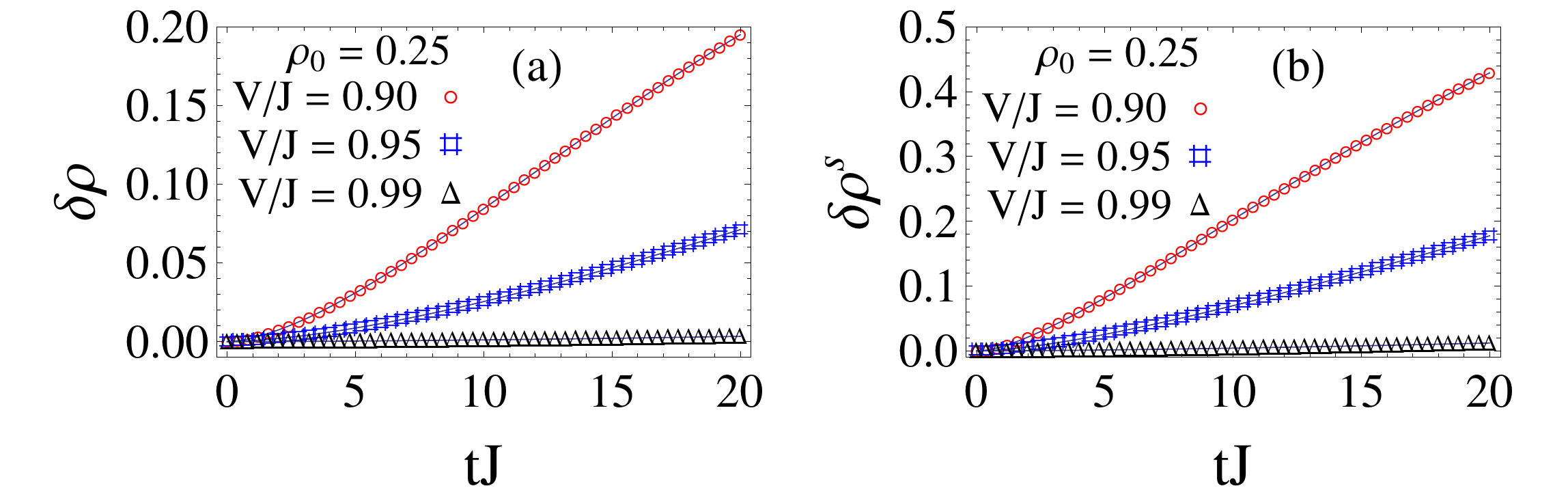}\\
\includegraphics[width=.53\textwidth]{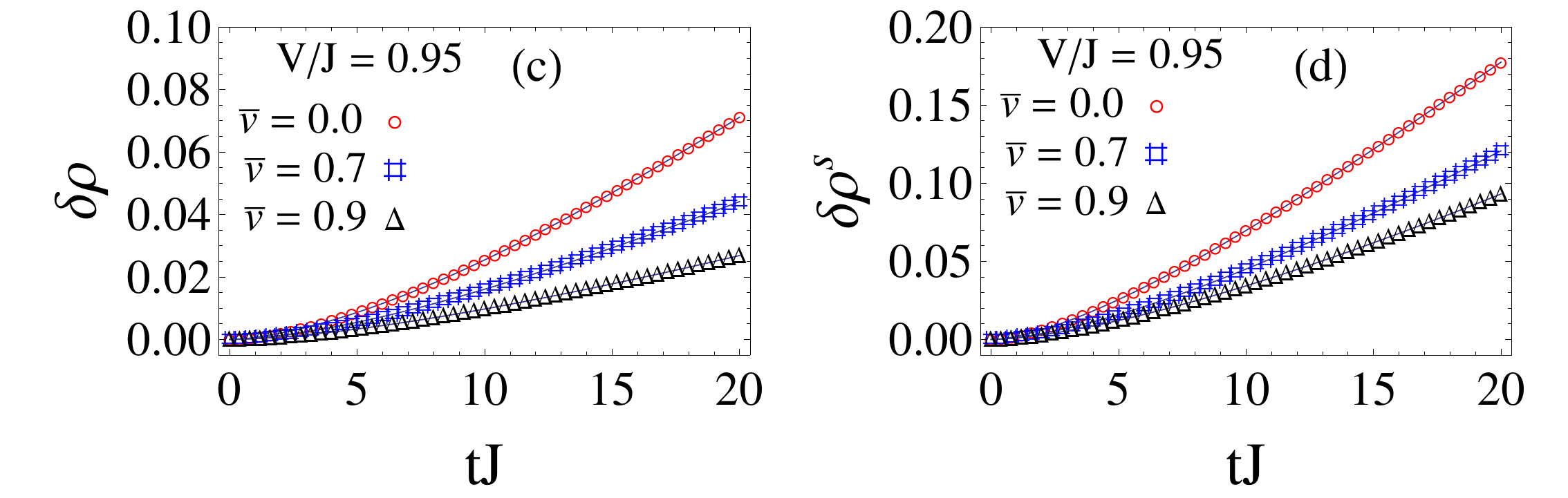}
\caption{(Color online) Differences using Eq.~\eqref{eq:difference_MFDMRG} between the DMRG results and the lattice mean-field results for the total density (left) and for the condensate density (right) for a stationary (top) and for moving (bottom) bright solitons ($\rho_0 = 0.25$) on a system with $L=40$ sites.}
\label{fig:fig2}
\end{figure}

To obtain a better measure for the life time of the solitons, we analyze in Fig.~\ref{fig:fig2} for the local observables (density $\rho$ and condensate density $\rho^s$, respectively) the discrepancy  between the t-DMRG evolution and the mean-field solution 
\begin{equation}
\delta \rho^{(s)} = \frac{\sqrt{\displaystyle\sum_i \left( \left\langle \rho_i^{(s)} \right\rangle_{\rm DMRG} - \left\langle \rho_i^{(s)} \right\rangle_{\rm MF} \right)^2}}{\sqrt{\displaystyle\sum_i  \left\langle \rho_i^{(s)} \right\rangle_{\rm MF}^2}}, 
\label{eq:difference_MFDMRG}
\end{equation}
similar to our analysis in Fig.~\ref{fig:compare_meanfield_lattice} which was based on Eq.~\eqref{eq:difference}. 
As shown in Fig.~\ref{fig:fig2}, $\delta \rho^{(s)}$ decreases significantly as $V/J$ approaches unity or as the speed of the soliton $\overline{v}$ (in units of the speed of sound) increases.   
This is associated to a widening of the initial density profile when increasing $V/J$ and a reduction of the peak amplitude for larger $\overline{v}$, so that we conclude from this analysis that for a variety of initial conditions the GPE and discrete-HGPE solitons can survive quantum fluctuations on the time scales treated. 
This is further confirmed in the following by the behavior of the entanglement entropy and the correlation functions. 

\subsection{Entanglement entropy and nearest neighbor correlations}
\label{sec:entropy}
\begin{figure}[t]
\includegraphics[width=.48\textwidth]{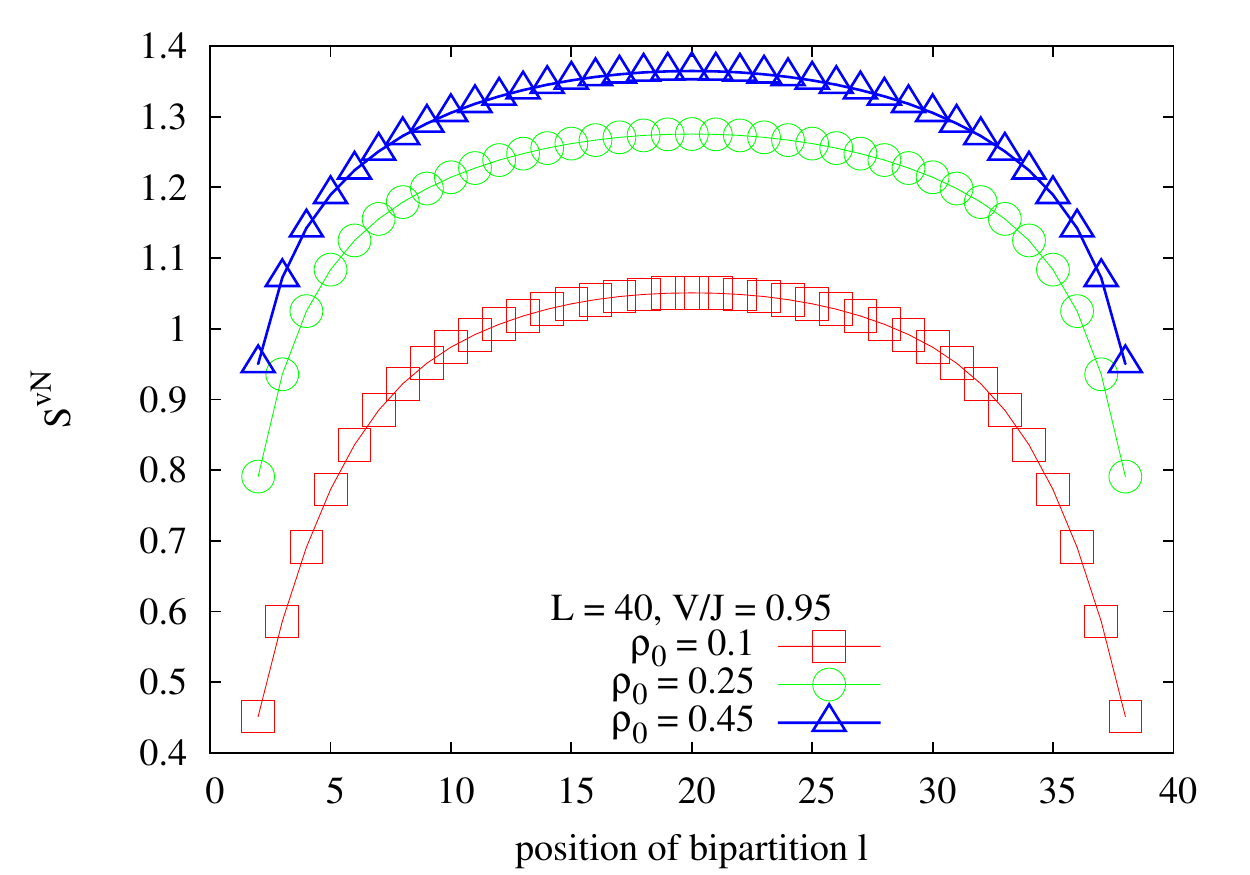}
\caption{(Color online) Entanglement entropy in ground states of the spin system Eq.~\eqref{eq:spinsystem} for $V/J = 0.95$, $L=40$ sites for values of $S^z_{\rm total}$ corresponding to the background density $\rho_0 = 0.1, \, 0.25,$ and $0.45$, respectively.}
\label{fig:entropy}
\end{figure}

A quantity that reveals the quantum nature of a state is the von Neumann or entanglement entropy in the system \cite{Amico} which is defined as 
\begin{equation}
S^{vN}_A = -{\rm Tr}(\rho_A {\rm log } \rho_A),
\end{equation}
with $\rho_A$ the reduced density matrix of a subsystem $A$ obtained by tracing out the degrees of freedom of the remaining part of the system $B$.  
From the Schmidt decomposition 
\begin{equation}
|\psi\rangle = \sum_i \sqrt{\lambda_i}|\phi_A^i \rangle |\phi_B^i \rangle
\label{eq:schmidt}
\end{equation}
it follows that
\begin{equation}
S^{vN} = -\sum_i \lambda_i \log \lambda_i,
\end{equation}
with $|\phi_A^i \rangle$ and $|\phi_B^i \rangle$ the eigenstates of the reduced density matrix of subsystem $A$ or $B$, respectively, and $\lambda_i$ the eigenvalues of the corresponding eigenstates. 
This quantity gives a measure for the entanglement between two subsystems. 
Since the initial states are product states on the lattice, $S^{vN}$ is exactly zero at the beginning of the time evolution since only one of the weights is finite with $\lambda_i = 1$ while the others are exactly zero. 
If $S^{vN}(t)$ remains zero (or very small) in the course of the time evolution, we conclude that quantum fluctuations do not strongly influence the nature of the initial product state, and so the value of $S^{vN}(t)$ gives an additional measure for the stability of the soliton solutions. 
Note that there are two variants of this analysis: in Refs.~\cite{MishmashPRL2009,Carr}, the entanglement entropy for a subsystem of one single site is measured with respect to the remainder of the system. 
However, within the DMRG framework it is easier to consider the time evolution of the entanglement entropy for all bipartitions of the system as it is automatically computed in the course of the DMRG procedure. 
For simplicity, and since it gives a similar measure for the stability of the soliton, we consider here the latter.  
In addition, the behavior of this quantity for ground states of finite spin chains is well known from conformal field theory \cite{Calabrese:2005p04010}, and the numerical values can be obtained easily from the DMRG.
This allows us to compare the values of the entanglement entropy during the time evolution to the ones of the strongly correlated ground state of the system which serves as a reference for how strongly  entangled the state has become during the time evolution. 
In Fig.~\ref{fig:entropy} we show typical result for the entanglement entropy in ground states of the spin system Eq.~\eqref{eq:spinsystem} with $L=40$ sites, $V/J=0.95$ and $S_z^{\rm total}$ corresponding to $\rho_0 = 0.1, \, 0.25$ and $0.45$, respectively.  
As can be seen, the numerical value in the center of the system increases with $\rho_0$ and reaches $S^{vN, \, center} \approx 1.35$ for $\rho_0 = 0.45$. 


A second estimate for the strength of the entanglement growth is to compare to the maximal possible entanglement entropy in a generic spin-$1/2$ chain with $L$ sites.  
Consider a bipartition of the chain into $M$ and $L-M$ spins with $M \le L-M$.   
Since the dimension of the Hilbert space of a chain of $M$ spins is $2^M$, a maximally entangled state is obtained when all $\lambda_j = 1/2^M$.
This state has hence an entropy 
\begin{equation*}
S^{vN,\, max} = -\sum_j \lambda_j \log \lambda_j = M \log 2.
\end{equation*}
For a system of $L=40$ sites and a bipartition $M=L/2$ we therefore obtain $S^{vN,\, max} \approx 13.86$, i.e. it is a factor of $\sim10$ larger than the one in the ground state for the same bipartition.


\begin{figure*}[t]
\includegraphics[width=.98\textwidth]{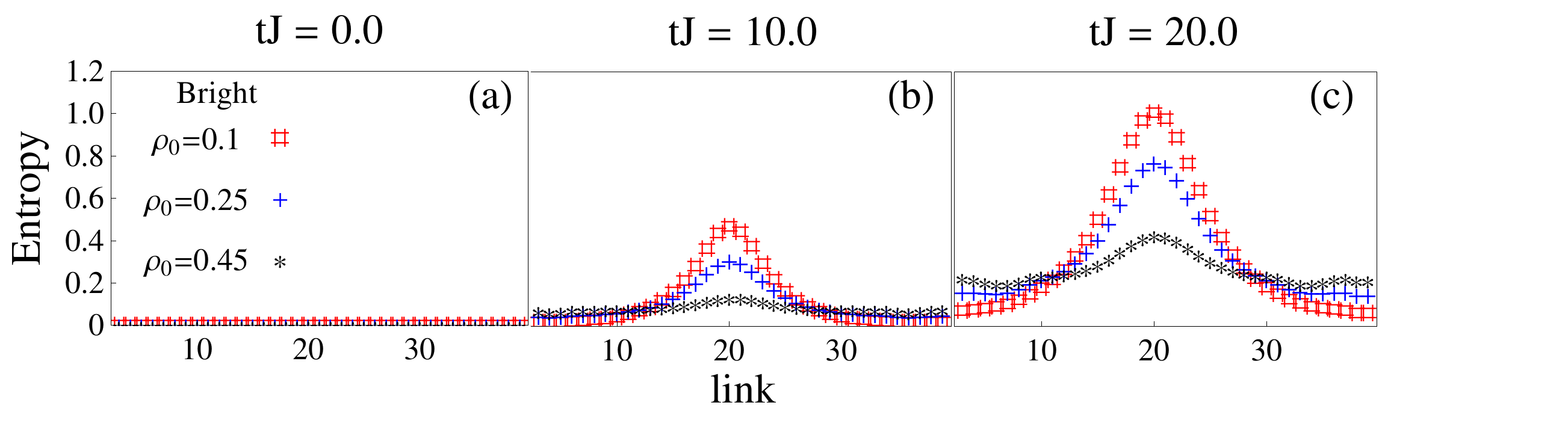} \\
\includegraphics[width=.98\textwidth]{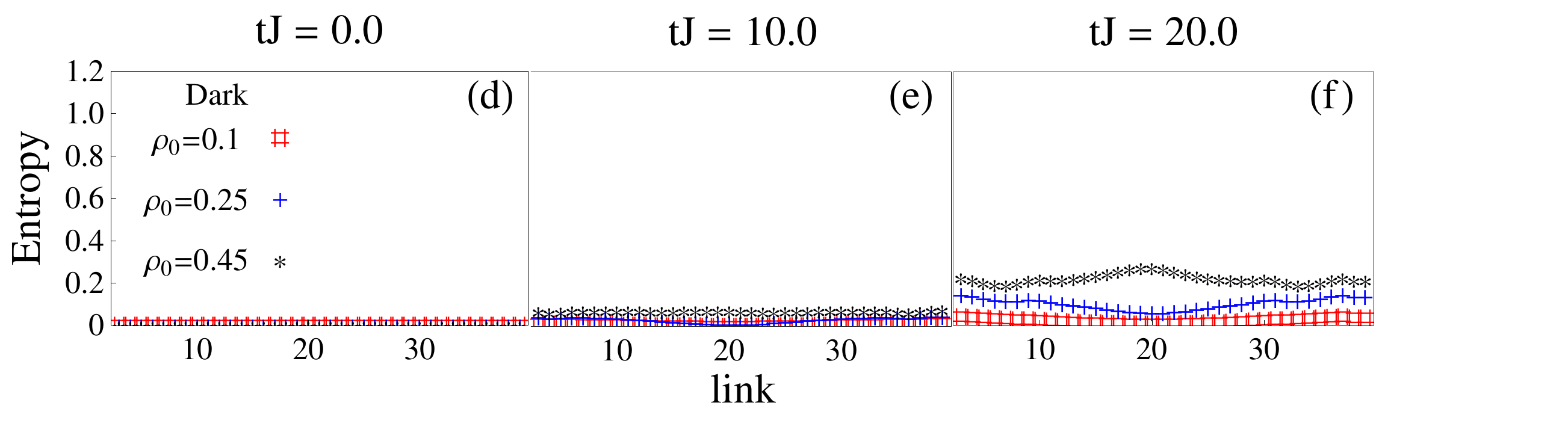}\\
\caption{(Color online) Entanglement entropies as a function of the subsystem size of the stationary bright soliton (top) and dark solitons (bottom) at different times for $V/J = 0.95$. 
}
\label{fig:fig3}
\end{figure*}

We now compare this values to the ones reached in the time evolution of the solitons.  
In Fig.~\ref{fig:fig3} we display the entanglement growth of both the dark and the bright soliton at $\rho_0 = 0.1, \, 0.25$ and $0.45$, respectively.    
We obtain that the entanglement growth is strongest at low fillings ($\rho_0 = 0.1$), and it is larger for the bright soliton than for the dark one. 
For $\rho_0 = 0.45$, the maximum value for the bright soliton is $S^{vN} \approx 0.4$, and for the dark soliton $S^{vN} \approx 0.25$ -- both values are significantly smaller than the one in the corresponding ground state, and much smaller than the one of the maximally entangled state. 
This shows that on the time scale treated, the state is significantly closer to a product state than to a strongly correlated ground state of the same system, or than to a maximally entangled state. 
Since the entanglement is not negligible, quantum fluctuations play an important role for the characterization of the state towards the end of the considered time evolution, but they are  not strong enough to fully destroy the product nature of the initial state. 

Note that the entanglement growth for the bright soliton for $\rho_0 = 0.1$ is significantly larger than for $\rho_0 = 0.45$. 
This is connected to the fact that also for the local observables the corresponding initial state decays much faster. 
The entanglement entropy can be used as a measure to compare the stability of the initial states at $\rho_0 = 0.1$ and $\rho_0 = 0.45$: it appears that the bright soliton at $\rho_0 = 0.1$ is about half as stable as the one at $\rho_0 = 0.45$. 
This is reflected in the numerical values of $\delta \rho^{(s)}(t)$ which also show  approximately a factor of two between the two cases. 

While the entanglement entropy at the center of the system shows a peak for the bright soliton, it possesses a minimum for the dark soliton. 
This can be understood by the fact that the dark soliton has fewer particles at the center of the system and so quantum fluctuations are less pronounced. 
Due to particle-hole symmetry, the dark and bright soliton evolutions for $\rho_0 \ge 1/2$ possess the same behavior.

\begin{figure}[t]
\includegraphics[width=.48\textwidth]{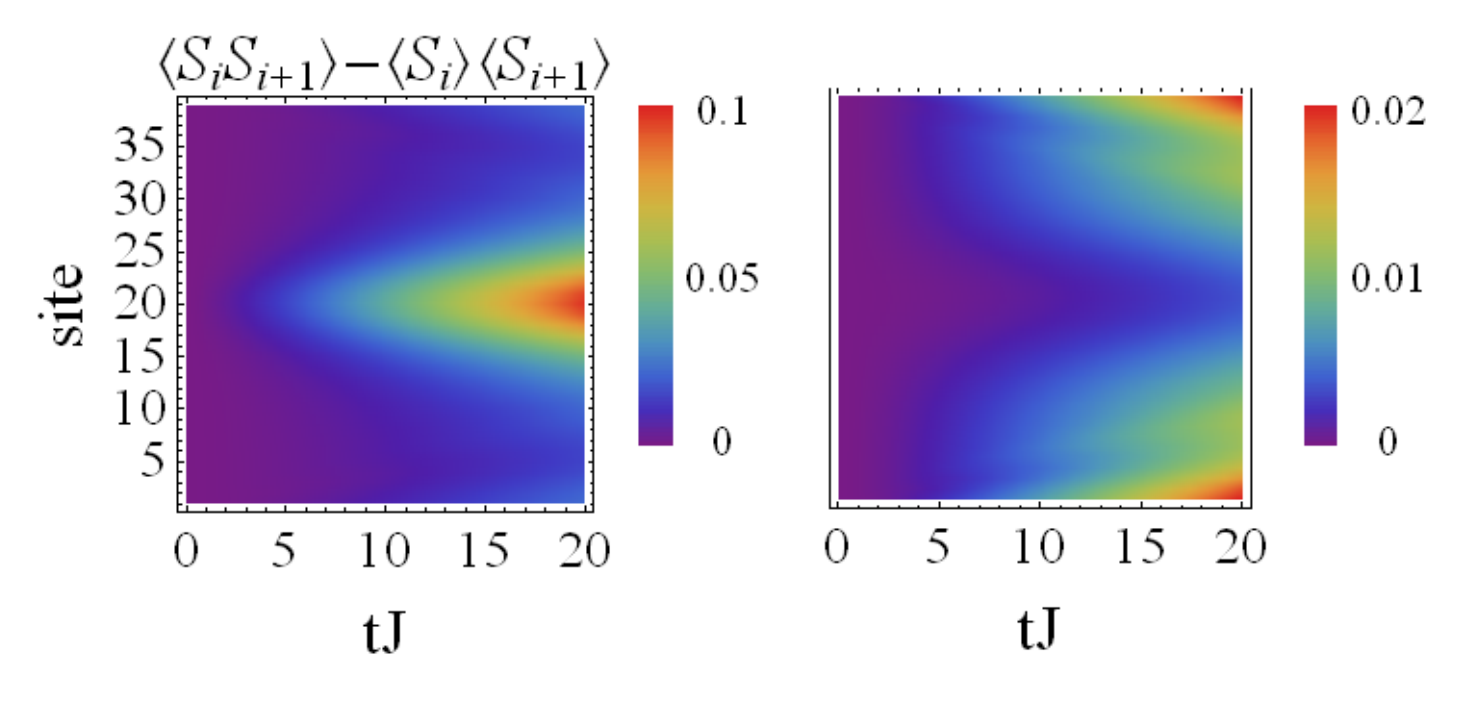} \\
\includegraphics[width=.48\textwidth]{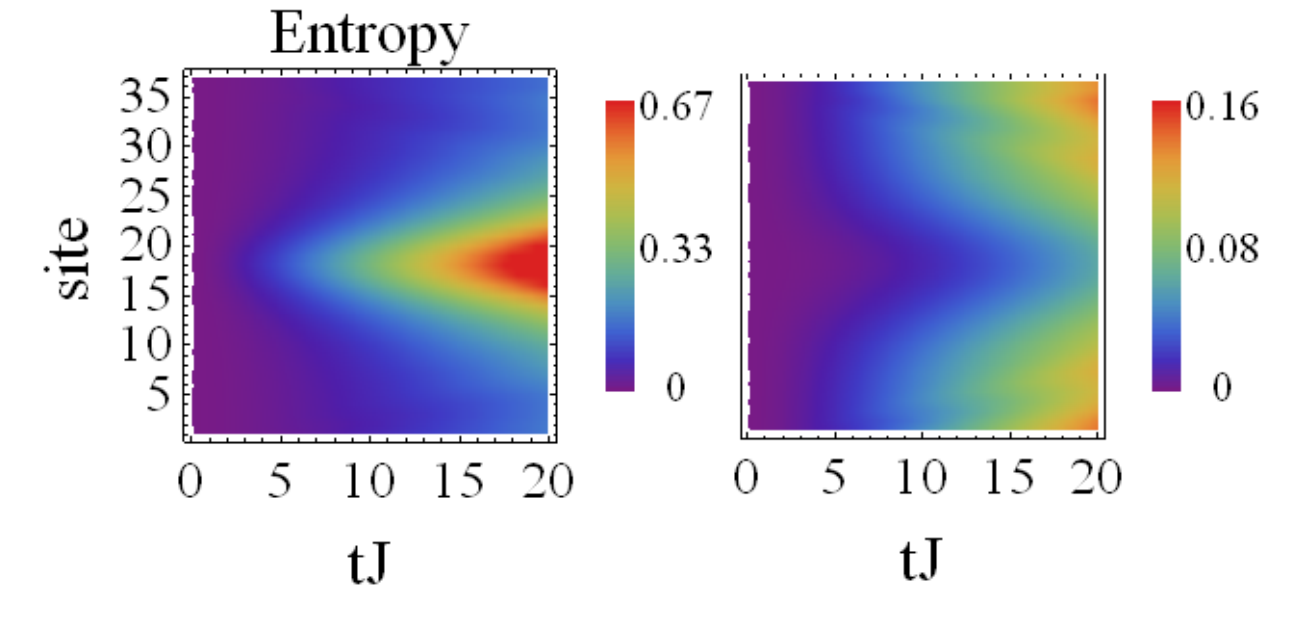}
\caption{(Color online) Top: contour plot of the time evolution of the nearest neighbor spin correlations $\langle S_i \cdot S_{i+1} \rangle - \langle S_i \rangle \cdot \langle S_{i+1} \rangle$ on the whole lattice for the stationary bright soliton (left) and the dark soliton (right) for $\rho_0 =0.25$ and $V/J = 0.95$. Bottom: time evolution of the entanglement entropy for the same parameters.}
\label{fig:fig4}
\end{figure}

The behavior of the entanglement entropy can be compared to the local spin fluctuations and correlations in the system. 
In the mean field approach, at all times the coherent spin state enforces that $\rho_i ^{s(MF)} = \rho_i(1-\rho_i)$. 
This relation can be expressed in terms of spin observables, leading to $\langle S^x_i \rangle^2 + \langle S^y_i \rangle^2 + \langle S^z_i \rangle^2 = 1/4$ on each site, realizing a constraint on the local spin fluctuations.  
In the full quantum dynamics, this constraint is broken, so that the initial coherent state becomes modified, and entanglement is induced in the system \cite{Toth:2009p042334}.     

\begin{figure}[t]
\includegraphics[width=.48\textwidth]{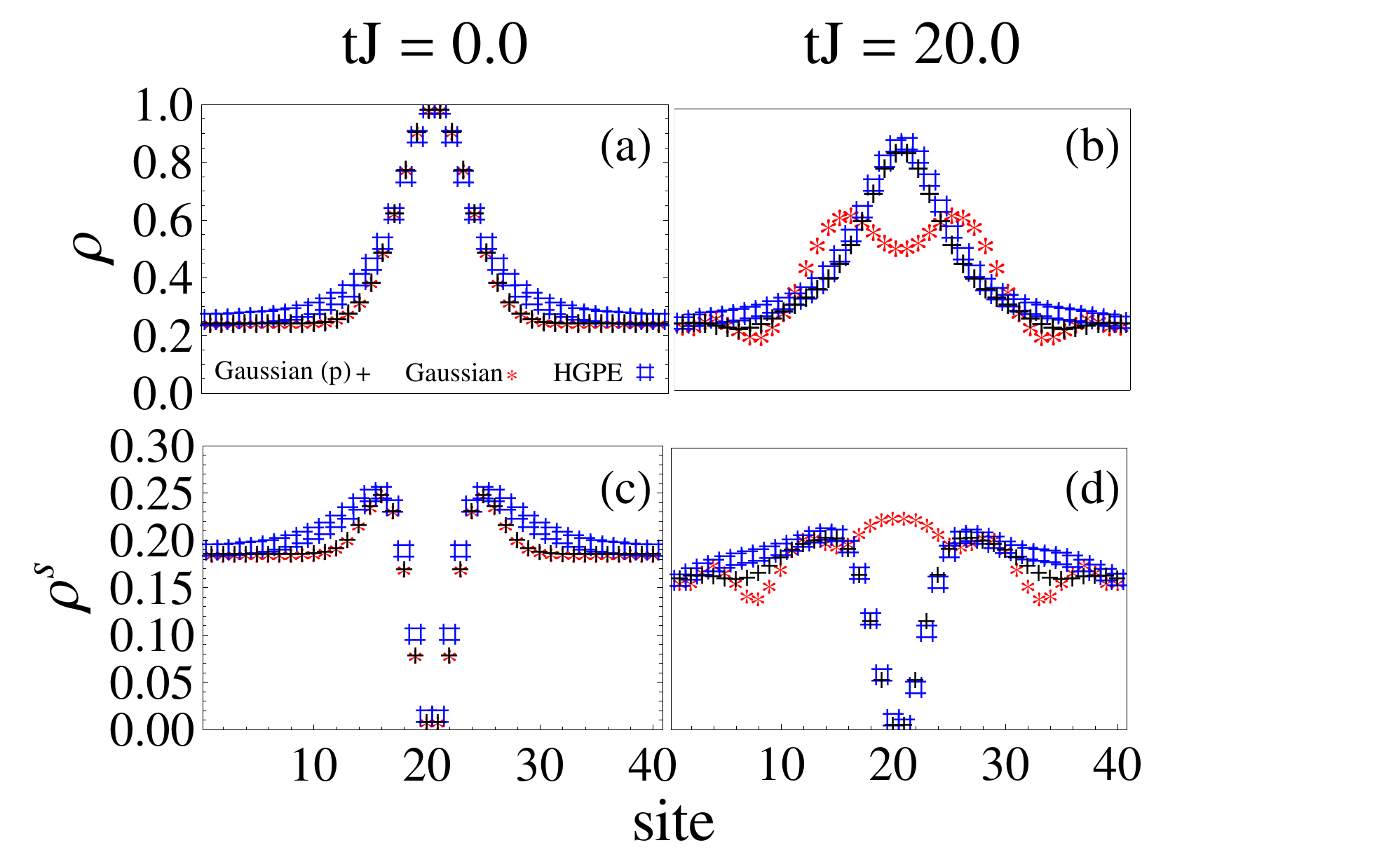}\\  
\includegraphics[width=.48\textwidth]{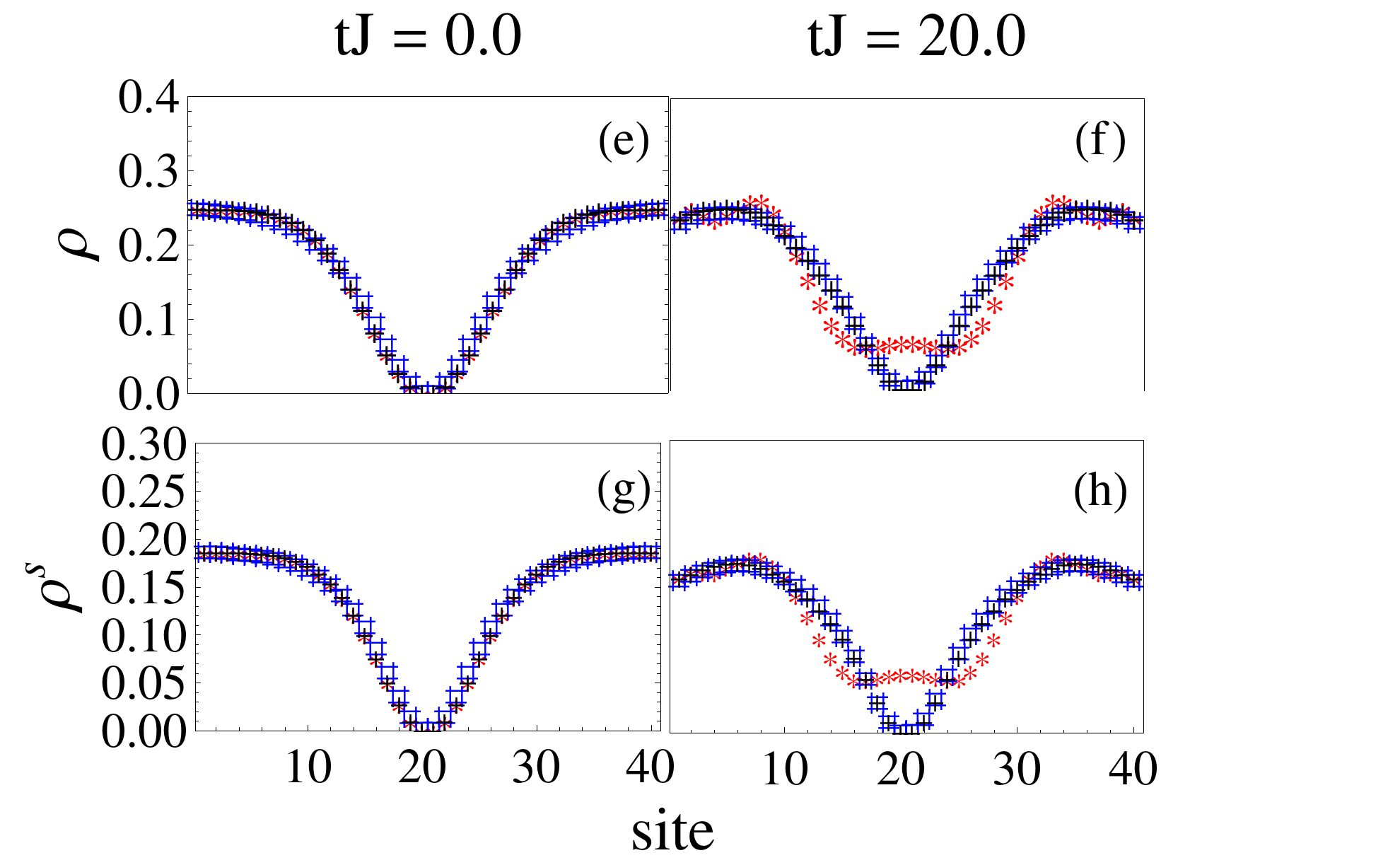}
\caption{(Color online) Comparison of the t-DMRG evolution of a bright (top) and dark (bottom) discrete HGPE soliton ($\#$) to the evolution of an initial Gaussian state with (+) and without $(*)$ phase imprinting for $\rho_0=0.25$ and $V/J = 0.95$.}
\label{fig:fig5}
\end{figure}

The entanglement entropy is related to the long distance correlations and has been extensively studied in spin systems \cite{Latorre:2009p504002,Alba:2009p10020,Vidal:2003p227902,Nienhuis:2009p02063,Sato:2007p8739}.  
It is therefore interesting to consider the growth of correlations in our system in the course of the time evolution. 
For simplicity, and since they are the most relevant ones for experiments, we consider nearest neighbor spin correlations $\langle S_i \cdot S_{i+1} \rangle - \langle S_i \rangle \cdot \langle S_{i+1} \rangle$.
The results shown in Fig.~\ref{fig:fig4} show similar behavior to the entropy dynamics.  

\subsection{Gaussian initial states}
\label{sec:gaussian}

In this section, we test the stability of the discrete HGPE soliton solutions to modifications of the initial state. 
Specifically, we compare the time evolution of these solitons to that of a Gaussian initial state (both obtained using the adaptive t-DMRG)
\begin{equation}
\rho(x) \sim e^{-\frac{x^2}{2\sigma^2}},
\end{equation}
which might be easier to implement in  experiments \cite{Reinhardt1997,Burger:1999p5198,Denschlag:2000p97}.  
We analyze the dynamics for initial states with and without a phase shift of $\pi$ across the center in order to compare the evolution of an initial state with a similar shape and phase properties as the HGPE soliton to one which has only a similar shape.   
As discussed in Sec.~\ref{sec:DMRG}, the initial state is created via a Gaussian external field.  

The obtained results are shown in Fig.~\ref{fig:fig5}. 
As can be seen, the Gaussian state with a phase jump remains stable and appears to be a very good approximation to the discrete HGPE soliton. 
In contrast, without the phase jump, the initial wave packet quickly disperses. 
Note that due to the lattice the wave packet can disperse by creating two peaks moving in opposite directions. 
This is due to the deviation of the $\cos(k)$ dispersion of the lattice from the dispersion $\sim k^2$ of a free particle and comes into appearance if the number of particles is high enough. 

We conclude, therefore, that once the phase jump is implemented, it is not necessary in the experiments to implement initial states which have exactly the form of the discrete HGPE solitons. 

\section{Experimental Realizations}
\label{sec:experiments}

In this section, we discuss possible realizations of the models introduced in Sec.~\ref{sec:model}. 
We start with the experimental implementation of the extended Bose Hubbard model, Eq.~\eqref{eq:BH}, and its fermionic variant.   
The nearest neighbor interaction term $V$ can be possibly generated in bosonic or spin polarized fermionic systems via long-range electric~\cite{GoralPRL} or magnetic~\cite{GriesmaierPRL} dipolar interactions  as discussed below or with a short-range interaction between atoms in higher bands of the lattice~\cite{ScarolaPRL}.   
The hard-core constraint for bosons requires increasing the interactions so that there is a large energy difference between states with a different number of bosons per site. 
This can be achieved by tuning the scattering length via a Feshbach resonance~\cite{Bloch_RMP}. 
Note that in this type of implementation, in which the  spin 1/2 degrees of freedom correspond to sites with zero and one atom, the {\it sign} of the $U$ and $V$ interaction is determined by the scattering length and thus is the same for both.  
In our proposal, we need attractive $V$ interaction, so also the Hubbard $U$ will be attractive.  
However, note that also in this case it is possible to realize the hard-core constraint: 
even though the states with one or zero atom per site do not belong to the ground state manifold,  when prepared, they are metastable since there is no way to dump the excess energy,  at least when prepared in the lowest band \cite{SensarmaPRL}. 
This is since the bandwidth in a lattice is finite, which is known to lead also to repulsively bound pairs \cite{WinklerNature}. 
In our case, however, it prevents double occupancies, which for $|U| \rightarrow \infty$ corresponds to the HCB limit. 
A possible way to proceed then is to prepare the ground state in the repulsive side of the Feshbach resonance and then quickly ramp the magnetic field to the attractive side in which the evolution takes place. 
The atoms now are still in the lowest band and need to be promoted to higher bands using, e.g.,  similar techniques to the ones discussed in Ref.~\onlinecite{MuellerPRL}.  
Note that the requirement of populating higher bands can indeed lead to an additional relaxation.  
In the fermionic system the decay to the lowest band can be blocked by filling the lowest band.  
The lifetime of bosons in higher bands on the other hand does require further investigation but at  least recent experiments in 2D \cite{MuellerPRL} reveal that it can be $10-100$ times longer than the characteristic time scale for intersite tunneling.  

In a recent proposal, it is shown that the XXZ spin model Eq.~(\ref{eq:spinsystem}) and the spinless fermion model Eq.~(\ref{eq:sfmodel}) both can be realized in systems of polar molecules on optical lattices, even though with a long-range $1/r^3$ decay of the interactions rather than nearest-neighbor interactions only.   
Two different paths allow the study of the soliton dynamics in such experiments: 
First, as discussed in detail in Refs.~\cite{gorshkov1,gorshkov2}, the spin model Eq.~(\ref{eq:spinsystem}) can be directly implemented in the case of unit filling (i.e., one molecule per site of the optical lattice) by selecting two rotational eigenstates of the molecules which emulate the two spin degrees of freedom of the $S=1/2$ chain.  
The parameters of the system can then be tuned via external DC electrical and microwave fields. 
The second implementation is by populating the lattice with molecules which are all in the same rotational eigenstate, emulating a spin polarized system. 
Since the dipolar interaction decays quickly, we presume that the effect of the interactions beyond nearest neighbor on the soliton dynamics should be very small, so that both realizations can be used to study the soliton dynamics.  
We leave a detailed study of the effect of the interaction terms beyond nearest-neighbor sites on the dynamics of the solitons to further studies. 

\section{ Summary}
\label{sec:summary}

We have analyzed the stability and lifetime of HGPE solitons on 1D lattice systems driven by a XXZ-Hamiltonian which can model the behavior of bosonic atoms, fermionic polar molecules, spin systems,  and spin-polarized itinerant fermions on optical lattices and in condensed matter systems.  
We compare the dynamics obtained in a mean field approximation to the full quantum evolution obtained using the adaptive t-DMRG and find that the solitons remain stable under the full quantum evolution on time scales $t \sim 20/J$, where $J/2$ is the unit of the hopping.    
This is quantified by the entanglement entropy which remains smaller than the one in the ground state of the corresponding spin system and significantly smaller then the one of a maximally entangled state on this time scale.  
Similar to the findings of Refs.~\cite{MishmashPRL2009,Carr}, for longer times the soliton decays. 
However, given the time scales reachable by ongoing experiments with optical lattices, this should suffice to identify this effect in the lab. 
In addition, we find that imperfections in the creation of the initial state should be of minor importance, as long as the density profile and the phase jump are similar to the ones of the proposed soliton solutions. 
This is exemplified by a Gaussian initial state, which in the case of a phase jump shows good agreement with the soliton solution, while in the absence of the phase jump becomes completely unstable. 
Due to the tunability of parameters either via Fesh\-bach resonances for atomic systems or via electric and microwave fields in the case of polar molecules, the possibility of realizing both, bright and dark solitons in strongly interacting systems, adds a new paradigm to the existence of coherent non-linear modes in systems of ultracold quantum gases.  



\section*{Acknowledgments}
We acknowledge financial support by 
ONR grant N00014-09-1-1025A, grant 70NANB7H6138 Am 001 by NIST, 
by NSF (PHYS 07-03278, PFC, PIF-0904017, and DMR-0955707), the AFOSR, and the ARO (DARPA-OLE). 
R.B. thanks the Department of Science and Technology, India, for financial support.

\end{document}